\input{aipcheck}

\documentclass[
    ,final            
  ]
  {aipproc}

\layoutstyle{8x11double}
\usepackage{amsmath,amssymb,exscale}

\newcommand{\be}{\begin{equation}}
\newcommand{\ee}{\end{equation}}
\newcommand{\ba}{\begin{eqnarray}}
\newcommand{\ea}{\end{eqnarray}}
\def\simlt{\mathrel{\lower2.5pt\vbox{\lineskip=0pt\baselineskip=0pt
             \hbox{$<$}\hbox{$\sim$}}}}
\def\simgt{\mathrel{\lower2.5pt\vbox{\lineskip=0pt\baselineskip=0pt
             \hbox{$>$}\hbox{$\sim$}}}}

\begin{document}

\title{Split Supersymmetry in String Theory}


\author{I.~Antoniadis\footnote{On leave from CPHT 
(UMR CNRS 7644) Ecole Polytechnique, F-91128 Palaiseau}}{
address={Department of Physics, CERN - Theory Division, 1211 Geneva 23, Switzerland}
}

\begin{abstract}
Type I string theory in the presence of internal magnetic fields 
provides a concrete realization of split supersymmetry. To lowest order,
gauginos are massless while squarks and sleptons are superheavy. 
For weak magnetic fields, the correct Standard Model spectrum guarantees 
gauge coupling unification with $\sin^2{\theta_W}=3/8$ at the compactification 
scale of $M_{\rm GUT}\simeq 2 \times 10^{16}$ GeV. I discuss mechanisms
for generating gaugino and higgsino masses at the TeV scale, as well as
generalizations to models with split extended supersymmetry in the gauge sector.
\end{abstract}

\maketitle


\section{Introduction}

During the last decades, physics beyond the Standard Model (SM) was guided
from the stabilization of mass hierarchy. For instance, compositeness, supersymmetry, 
extra dimensions, low string scale and little Higgs are different approaches to
address the hierarchy. However, the actual precision tests, implying the 
absence of any deviation from the SM to a great accuracy, suggest that
any new physics at a TeV needs to be fine-tuned at the per-cent level.
Thus, either the underlying theory beyond the SM is very special, or 
our notion of naturalness should be reconsidered. The latter is also motivated
from the recent evidence for the presence of a tiny non-vanishing cosmological
constant that raises another more severe hierarchy problem. 
This raises the possibility that the same mechanism may solve both
problems and casts some doubts on all previous proposals.

On the other hand, the necessity of a Dark Matter (DM) candidate and 
the fact that LEP data favor the unification of the three SM gauge couplings 
are smoking guns for the presence of new physics at high energies.
Supersymmetry is then a nice candidate offering both properties.
Moreover, it arises naturally in string theory, which provides a framework
for incorporating the gravitational interaction in our quantum picture
of the universe. It was then proposed to consider that supersymmetry
might be broken at high energies without solving the gauge hierarchy
problem. More precisely, making squarks and sleptons heavy does not
spoil unification and the existence of a DM candidate while at the
same time it gets rid of all unwanted features of the supersymmetric
SM related to its complicated scalar sector. On the other hand,
experimental hints to the existence of supersymmetry persist since
there are still gauginos and higgsinos at the electroweak scale. 
This is the so-called split supersymmetry 
framework~\cite{Arkani-Hamed:2004fb,Giudice:2004tc}.

Split supersymmetry has a natural realization in type I string theory
with magnetized D9-branes, or equivalently with branes at 
angles~\cite{Antoniadis:2004dt}. We first show that the general spectrum
has the required properties and then discuss the conditions for gauge
coupling unification near the string scale. It turns out that equality of the
two non-abelian couplings is a consequence of the correct SM spectrum
for weak magnetic fields, while the value for the weak angle 
$\sin^2\theta_W=3/8$ is easily obtained even in simple constructions.
Indeed, we perform a general study of SM embedding in three
brane stacks and find a simple model realizing the conditions for
unification~\cite{Antoniadis:2004dt}. We then discuss mass scales 
and in particular a mechanism generating light gaugino and higgsino 
masses in the TeV region, while scalars are superheavy, of order 
$10^{13}$ GeV~\cite{Antoniadis:2005sd}.
Finally, we show how splitting supersymmetry reconciles toroidal
models of intersecting branes with unification~\cite{Antoniadis:2005em}. 
The gauge sector in these models arises in multiplets of extended 
supersymmetry while matter states are in ${\cal N}=1$ representations.
In general, split supersymmetry offers new possibilities for realistic
string model building, that were previously unavailable because
they were mainly restricted in the context of large dimensions and low 
string scale~\cite{ld,Arkani-Hamed:1998rs}.

\section{General framework}

We start with type I string theory, or equivalently type IIB with 
orientifold 9-planes and D9-branes~\cite{Angelantonj:2002ct}. 
Upon compactification in four dimensions on a Calabi-Yau manifold, 
one gets ${\cal N}=2$ supersymmetry in the bulk and 
${\cal N}=1$ on the branes. Moreover, various fluxes can be turned on, to 
stabilize part or all of the closed string moduli. We then turn on 
internal magnetic fields~\cite{Bachas:1995ik, Angelantonj:2000hi}, 
which, in the T-dual picture, amounts to 
intersecting branes~\cite{Berkooz:1996km, bi}. 
For generic angles, or equivalently for 
arbitrary magnetic fields, supersymmetry is spontaneously broken and 
described by effective D-terms in the four-dimensional (4d) 
theory~\cite{Bachas:1995ik}. In the weak field limit, 
$|H|\alpha'<1$ with $\alpha'$ the string Regge slope, 
the resulting mass shifts are given by:
\be
\delta M^2=(2k+1)|qH|+2qH\Sigma\quad ;\quad k=0,1,2,\dots\, ,
\label{deltam}
\ee
where $H$ is the magnetic field of an abelian 
gauge symmetry, corresponding to a  Cartan generator of the higher 
dimensional gauge group, on a non-contractible 2-cycle of the 
internal manifold. $\Sigma$ is the corresponding projection of the 
spin operator, $k$ is the Landau level and  $q=q_L+q_R$ is the charge 
of the state, given by the sum of the left and right charges of the 
endpoints of the associated open string. We recall that the exact 
string mass formula has the same form as (\ref{deltam}) with $qH$ 
replaced by:
\be
qH\longrightarrow\theta_L+\theta_R\qquad ;\qquad 
\theta_{L,R}=\arctan(q_{L,R}H\alpha')\, .
\label{stringdeltam}
\ee
Obviously, the field theory expression 
(\ref{deltam}) is reproduced in the weak field limit.

The Gauss law for the 
magnetic flux implies that the field $H$ is quantized in terms of 
the area of the corresponding 2-cycle $A$:
\be
H={m\over nA}\, ,
\label{Hquant}
\ee
where the integers $m,n$ correspond 
to the respective magnetic and electric charges; $m$ is the 
quantized flux and $n$ is the wrapping number of the higher 
dimensional brane around the corresponding internal 2-cycle. 

For simplicity, we consider below the case where 
the internal manifold is a product of three factorized tori 
$\prod_{I=1}^3 T^2_{(I)}$. Then, the mass formula (\ref{deltam})
becomes:
\be
\delta M^2=\sum_I(2k_I+1)|qH_I|+2qH_I\Sigma_I\, ,
\label{deltamI}
\ee
where $\Sigma_I$ is the projection of 
the internal helicity along the $I$-th plane. For a ten-dimensional (10d) spinor,
its eigenvalues are $\Sigma_I=\pm 1/2$, while for a 10d 
vector $\Sigma_I=\pm 1$ in one of the planes $I=I_0$
and zero in the other two $(I\ne I_0)$. Thus, charged higher dimensional
scalars become massive, fermions lead to chiral 4d zero modes if all 
$H_I\ne 0$, while the lightest scalars coming from 10d vectors have masses
\be
M_0^2=\left\{
\begin{matrix}
|qH_1|+|qH_2|-|qH_3|\cr
|qH_1|-|qH_2|+|qH_3|\cr
-|qH_1|+|qH_2|+|qH_3|\cr
\end{matrix}\, .
\right.
\label{scalars}
\ee
Note that all of them can be made positive definite, avoiding
the Nielsen-Olesen instability, if all $H_I\ne 0$. Moreover, one can 
easily show that if a scalar mass vanishes, some supersymmetry 
remains unbroken~\cite{Angelantonj:2000hi,Berkooz:1996km}.

\section{Generic spectrum}

We turn on now 
several abelian magnetic fields $H_I^a$ of different Cartan generators 
$U(1)_a$, so that the gauge group is a product of unitary factors 
$\prod_a U(N_a)$ with $U(N_a)=SU(N_a)\times U(1)_a$. In an 
appropriate T-dual 
representation, it amounts to consider several stacks of 
D6-branes intersecting in the three internal tori at angles. An 
open string with one end on the $a$-th stack has charge $\pm 1$ under 
the $U(1)_a$, depending on its orientation, and is neutral with 
respect to all others. Using the results described above, the 
massless spectrum of the theory falls into three sectors~\cite{bi,Angelantonj:2000hi}:
\begin{enumerate}
\item Neutral open strings ending on 
the same stack, giving rise to ${\cal N}=1$ gauge supermultiplets of gauge 
bosons and gauginos.
\item Doubled charged open strings from a 
single stack, with charges $\pm 2$ under the corresponding $U(1)$, 
giving rise to massless fermions transforming in the antisymmetric or 
symmetric representation of the associated $SU(N)$ factor. Their 
bosonic superpartners become massive. The multiplicities of chiral fermions 
are given by:
\ba
{\rm Antisymmetric}\!\!\!\!\!\! &:&\!\!\!\!\!\! {1\over 2}\left(\prod_I 
2m_I^a\right)\left(\prod_J n_J^a+1\right)\nonumber\\
{\rm Symmetric}\!\!\!\!\!\! &:&\!\!\!\!\!\! {1\over 2}
\left(\prod_I 2m_I^a\right)\left(\prod_J n_J^a-1\right)
\label{dcmult}
\ea
where $m_I^a, 
n_I^a$ are the integers entering in the expression of the magnetic 
field (\ref{Hquant}). For orbifolds or more general Calabi-Yau spaces, 
the above multiplicities may be further reduced by the corresponding 
supersymmetry projection down to ${\cal N}=1$.

In the degenerate case where 
a magnetic field vanishes, say, along one of the tori ($m_I^a=0$ for 
some $I$), there are no chiral fermions in $d=4$ dimensions, but the 
same formula with the products extending over the other two 
magnetized tori gives the multiplicities of chiral fermions in $d=6$. 
In this case, chirality in four dimensions may arise only when the 
last $T^2$ compactification is combined with some additional 
orbifold-type projection.
\item Open strings stretched between 
two different brane stacks, with charges $\pm 1$ under each of the 
corresponding $U(1)$'s. They give rise to chiral fermions transforming 
in the bifundamental representation of the two associated unitary 
group factors. Their multiplicities, for toroidal compactifications, 
are given by:
\ba
(N_a,N_b)&:&\prod_I (m_I^a n_I^b+n_I^a 
m_I^b)
\nonumber\\
(N_a,{\overline N}_b)&:&\prod_I (m_I^a n_I^b-n_I^a 
m_I^b)\, .
\label{scmult}
\ea
As in the previous case, when a factor in the products of the above 
multiplicities vanishes, there are no 4d chiral 
fermions, but the same formula with the product restricted over the 
other two magnetized tori gives the corresponding multiplicity of 
chiral fermions in $d=6$.
\end{enumerate}

As mentioned already above, 
all charged bosons are massive. Massless scalars can appear only when 
some supersymmetry remains unbroken. 
It is now clear that this framework leads 
to models with a tree-level spectrum realizing the idea of split 
supersymmetry. Embedding the Standard Model (SM) in an appropriate 
configuration of D-brane stacks, one obtains tree-level
massless gauginos while 
all scalar superpartners of quarks and leptons typically get masses at the  scale 
of the magnetic fields, whose magnitude is set by the compactification scale 
of the corresponding internal space. 
On the other hand, the condition to obtain a (tree-level)
massless Higgs in the spectrum implies that 
supersymmetry remains unbroken in the Higgs sector, leading to a pair 
of massless higgsinos, as required by anomaly cancellation.

\section{Gauge coupling unification}

On general grounds, there are two conditions to obtain unification of
SM gauge interactions, consistently with extrapolation of
gauge couplings from low-energy data using the minimal supersymmetric
SM spectrum. (i) Equality of the $SU(3)$ color and weak $SU(2)$ 
non-abelian gauge couplings and (ii) the correct prediction for the weak 
mixing angle $\sin^2\theta_W=3/8$ at the grand unification (GUT) scale.
On the other hand, a generic D-brane model using several stacks, as
described in the framework of the previous section, does not satisfy
either of the two conditions. Indeed, this framework was developed
in connection to the idea of low-scale 
strings~\cite{Arkani-Hamed:1998rs}, where the concept of
unification is radically different from conventional GUTs.
In this section, we study precisely the general requirements for 
satisfying the first of the above two conditions, namely
natural unification of non-abelian gauge couplings.

The 4d non-abelian gauge coupling $\alpha_{N_a}$ of the 
$a$-th brane stack is given by:
\ba
{1\over \alpha_{N_a}}={V\over g_s}
\prod_I |n_I^a|\sqrt{1+(H_I^a\alpha')^2}\, ,
\label{ga}
\ea
where $g_s$ is the string coupling and $V$ the compactification volume 
in string units. 
The presence of the wrapping numbers $|n_I^a|$ can be understood from 
the fact that $|n_I^a|V_I$ is the effective area of the 2-torus $T^2_{(I)}$
wrapped $n_I^a$ times by the D9-brane, and $V=\prod_IV_I$.
The additional factor in the square root follows from the non-linear
Dirac-Born-Infeld (DBI) action of the abelian gauge field, 
$\sqrt{\det (\delta_{ij}+F_{ij}\alpha')}$, which in the case of two dimensions
with $F_{ij}=\epsilon_{ij}H$, it is reduced to $\sqrt{1+(H\alpha')^2}$.
Obviously, the expression (\ref{ga}) holds at the compactification scale, 
since above it gauge couplings receive 
important corrections and become higher dimensional. 
Finally, the gauge couplings of the associated abelian factors, in our 
convention of $U(1)$ charges, are given by
\be 
\alpha_{_{U(1)_a}}={\alpha_{N_a}\over{2N_a}}\, . 
\label{gua}
\ee
Here, non-abelian generators are normalized according to 
${\rm Tr}T^aT^b=\delta^{ab}/2$. 

From equation (\ref{ga}), it follows that 
unification of non-abelian gauge couplings holds if (i) 
$\prod_I|n_I^a|$ are independent of $a$, and (ii) the 
magnetic fields are either $a$-independent as well, or they are much 
smaller than the string scale. 

Condition (i) follows from eq.~(\ref{dcmult}), by 
requiring the absence of chiral fermions transforming in the 
symmetric representations of the non-abelian groups, {\em i.e.} no 
chiral $SU(3)$ color sextets and no weak $SU(2)$ 
triplets.

Condition (ii) of weak 
magnetic fields is more quantitative. Allowing for $1\%$ error in the 
unification condition at high scale, one should have 
$|H_I^a|\alpha'\simlt 0.1$. From the quantization condition 
(\ref{Hquant}), this implies that the volume $V\simgt 10^3$ for three 
magnetized tori, which is rather high to keep the theory weakly 
coupled above the compactification scale. Indeed, eq.~(\ref{ga})
gives a string coupling $g_s$ of order ${\cal O}(10)$ for gauge couplings
$\alpha_{N_a}\simeq 1/25$ at the unification scale. On the other hand, 
for one or two magnetized tori one obtains $V\simgt 10-10^2$, 
which is compatible with a string weak 
coupling regime $(g_s\sim 0.1-1)$. 
Fortunately, this condition 
can be partly relaxed in some direction, by requiring the absence of 
chiral antiquark doublets in the spectrum. Indeed eq.~(\ref{scmult}),
for open strings stretched between the strong $SU(3)$ and 
weak $SU(2)$ interactions brane stacks, 
implies the vanishing of one of the factors in the product. 
This leads to the equality of the ratio $m_I^a/n_I^a$ for the 
two stacks and for some $I$, and thus, to the equality of the two 
corresponding magnetic fields via eq.~(\ref{Hquant}).\footnote{This 
argument is true only when the $U(1)$ accompanying the weak 
interactions brane stack participates in the hypercharge combination. 
Otherwise, quark anti-doublets are equivalent to quark doublets.} As a result, 
the condition of perturbativity is weakened and becomes 
possible even in the case of three factorized magnetized tori.

The above analysis concerns the non-abelian couplings 
$\alpha_3$ and $\alpha_2$ of strong and weak interactions. 
The case of hypercharge is more subtle since 
it can be in general a linear combination of several $U(1)$'s coming 
from different brane stacks. In the following section, 
we present an explicit example with the correct prediction of the
weak mixing angle. It is based on a minimal SM 
embedding in three brane stacks with the hypercharge being a 
linear combination of two abelian factors. This provides an existence
proof that can be generalized in different constructions.
We notice for instance that in a class of supersymmetric
models with four brane stacks, the equality of the two non-abelian
couplings $\alpha_2=\alpha_3$ implies the value $3/8$ for
$\sin^2\theta_W$ at the unification scale~\cite{Blumenhagen:2003jy}.

\section{Minimal Standard Model embedding}

In this section, we perform a general study of SM embedding
in three brane stacks with gauge group $U(3)\times U(2)\times U(1)$~\cite{ar},
and present an explicit example having
realistic particle content and satisfying gauge coupling unification.

\begin{figure}[h]
\includegraphics[height=.2\textheight]{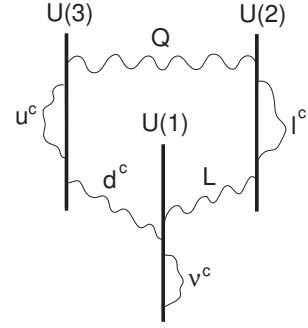}\\
\label{fig_modelA}
\caption{Pictorial representation of model A}
\end{figure}

\begin{figure}[h]
\includegraphics[height=.2\textheight]{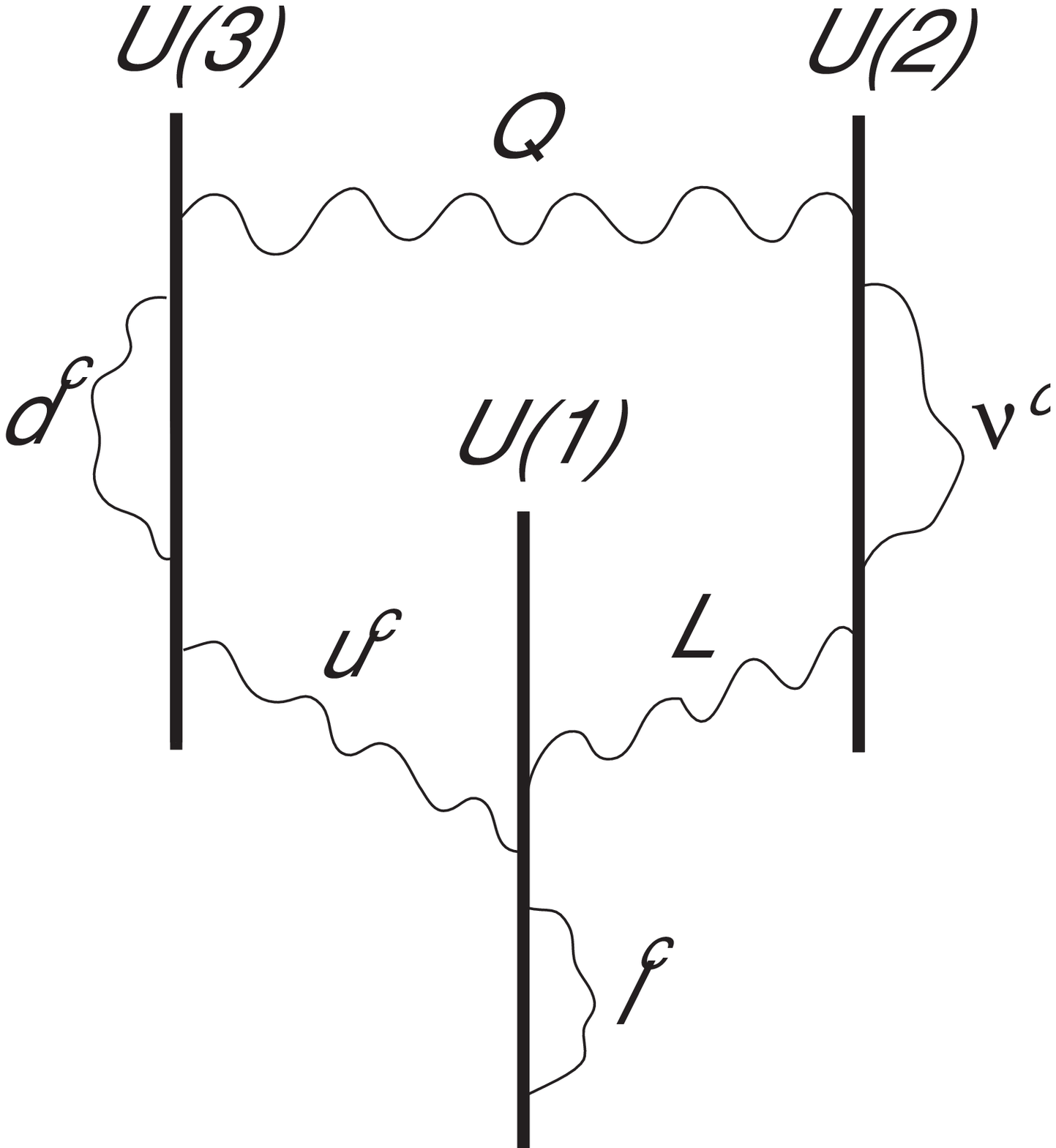}
\includegraphics[height=.2\textheight]{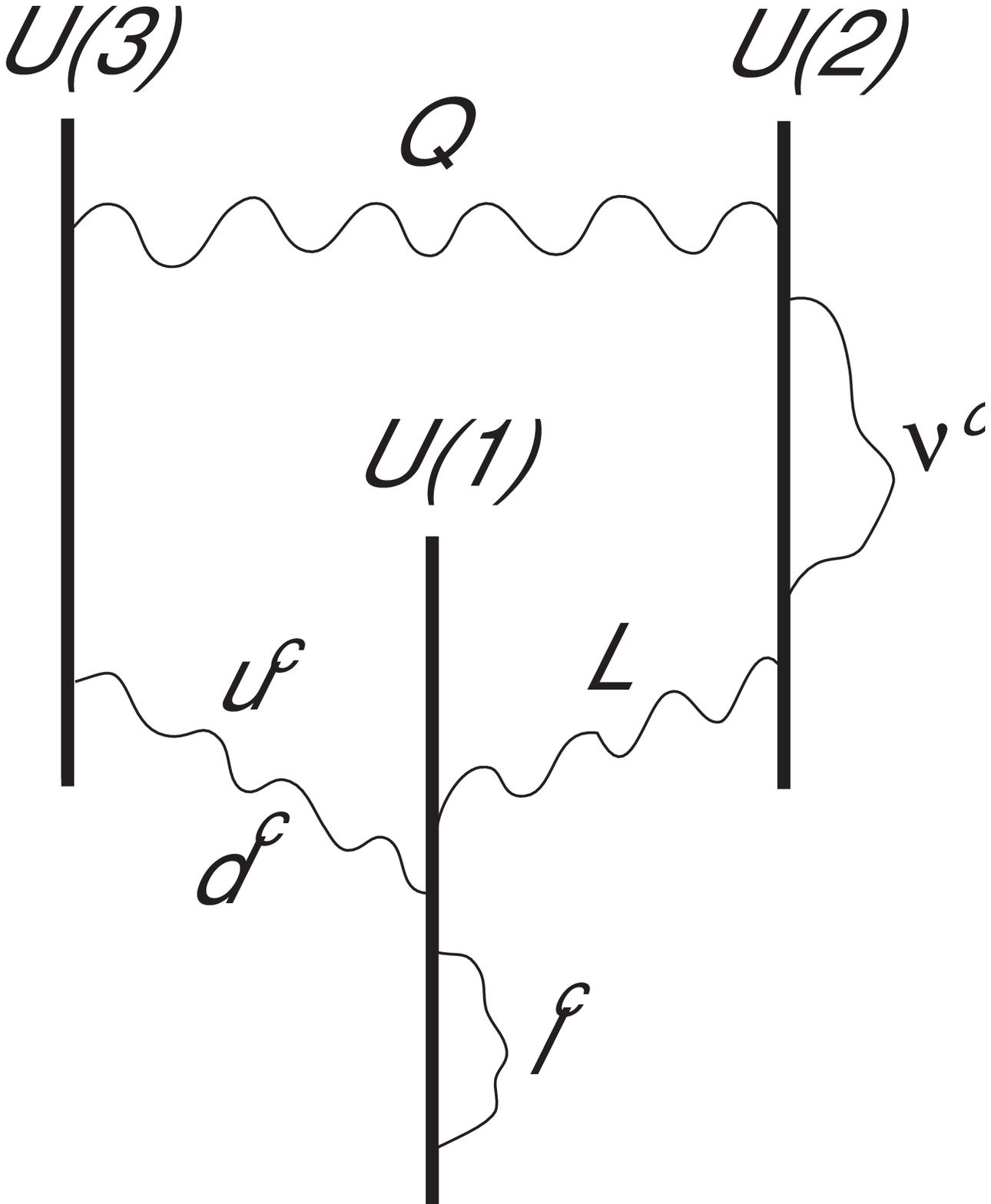}
\label{fig_modelB}
\caption{Pictorial representation of models B and C}
\end{figure}

The quark and lepton doublets ($Q$ and $L$) correspond 
to open strings stretched between the weak and the color or $U(1)$ 
branes, respectively. On the other hand, the $u^c$ and $d^c$ antiquarks 
can come from strings that are
either stretched between the color and $U(1)$ branes, or that have 
both ends on the color branes and transform in the antisymmetric 
representation of $U(3)$ (which is an anti-triplet). There are 
therefore three possible models, depending on whether it is the $u^c$ 
(model A), or the $d^c$ (model B), or none of them (model C), the 
state coming from the antisymmetric representation of color branes. 
It follows that the antilepton $l^c$ comes in a similar way  from 
open strings with both ends either on the weak brane stack and 
transforming in the antisymmetric representation of $U(2)$ which is 
an $SU(2)$ singlet (in model A), or on the abelian brane 
and transforming in the ``symmetric" representation of $U(1)$
(in models B and C). The three models are presented pictorially 
in Fig.~\ref{fig_modelA} (model A)
and Fig.~\ref{fig_modelB} (models B,C).


Thus, the members of a family of quarks and leptons have the 
following quantum numbers:
\ba
&&{\rm Model\ A}\nonumber\\
\!\!\!\!\!\!\!\! Q && ({\bf 3},{\bf 2};1,1,0)_{1/6}\nonumber\\
\!\!\!\!\!\!\!\! u^c && (\bar{\bf 3},{\bf 1};2,0,0)_{-2/3}\nonumber\\
\!\!\!\!\!\!\!\! d^c && (\bar{\bf 3},{\bf 1};-1,0,\varepsilon_d )_{1/3}
\label{modelA}\\
\!\!\!\!\!\!\!\! L && ({\bf 1},{\bf 2};0,-1,\varepsilon_L)_{-1/2}\nonumber\\
\!\!\!\!\!\!\!\! l^c && ({\bf 1},{\bf 1};0,2,0)_1\nonumber\\
\!\!\!\!\!\!\!\! \nu^c && ({\bf 1},{\bf 1};0,0,2\varepsilon_\nu)_0\nonumber
\ea
\ba
&&{\rm Model\ B}
\qquad\qquad\quad\ {\rm Model\ C}\nonumber\\
\!\!\!\!\!\!\!\! Q && ({\bf 3},{\bf 2};1,\varepsilon_Q,0)_{1/6}
\qquad\, ({\bf 3},{\bf 2};1,\varepsilon_Q,0)_{1/6}\nonumber\\
\!\!\!\!\!\!\!\! u^c && (\bar{\bf 3},{\bf 1};-1,0,1)_{-2/3}
\quad\ (\bar{\bf 3},{\bf 1};-1,0,1)_{-2/3}\nonumber\\
\!\!\!\!\!\!\!\! d^c && (\bar{\bf 3},{\bf 1};2,0,0)_{1/3}
\qquad\ \ \, (\bar{\bf 3},{\bf 1};-1,0,-1)_{1/3}
\label{modelB}\\
\!\!\!\!\!\!\!\! L && ({\bf 1},{\bf 2};0,\varepsilon_L,1)_{-1/2}
\quad\ \ \, ({\bf 1},{\bf 2};0,\varepsilon_L,1)_{-1/2}\nonumber\\
\!\!\!\!\!\!\!\! l^c && ({\bf 1},{\bf 1};0,0,-2)_1
\qquad\ \ ({\bf 1},{\bf 1};0,0,-2)_1\nonumber\\
\!\!\!\!\!\!\!\! \nu^c && ({\bf 1},{\bf 1};0,2\varepsilon_\nu,0)_0
\qquad\ \, ({\bf 1},{\bf 1};0,2\varepsilon_\nu,0)_0\nonumber
\ea
where the last three digits after the semi-column 
in the brackets are the charges under the three 
abelian factors $U(1)_3\times U(1)_2\times U(1)$, that we will call 
$Q_3$, $Q_2$ and $Q_1$ in the following, while the subscripts denote 
the corresponding hypercharges. The various sign ambiguities 
$\varepsilon_i=\pm 1$ are due to the fact that the corresponding 
abelian factor does not participate in the hypercharge combination 
(see below). 
In the last lines, we also give the quantum numbers of a possible 
right-handed neutrino in each of the three models. These are in fact 
all possible ways of embedding the SM spectrum in three sets of 
branes.

The hypercharge combination is:
\ba
\label{hyper}
{\rm Model\ A}\quad\ &:&\quad Y=-{1\over 3}Q_3+{1\over 2}Q_2\\
{\rm Model\ B, C}&:&\quad 
Y=\ \ \, {1\over 6}Q_3-{1\over 2}Q_1\nonumber
\ea
leading to 
the following expressions for the weak angle:
\ba
{\rm Model\ A}
&:&\sin^2\theta_W={1\over 2+2\alpha_2/3\alpha_3}
={3\over 8}\, {\bigg|}_{\alpha_{_2}=\alpha_{_3}}\\
{\rm Model\ B, C}&:&
\sin^2\theta_W={1\over 1+\alpha_2/2\alpha_1+\alpha_2/6\alpha_3}\nonumber\\
& &\qquad\quad\ ={6\over 7+3\alpha_2/\alpha_1}\, 
{\bigg|}_{\alpha_{_2}=\alpha_{_3}}\nonumber
\label{sintheta}
\ea
In the second part of the above equalities, we used the unification relation 
$\alpha_2=\alpha_3$, that can be naturally imposed as described in 
the previous section. It follows that model A admits natural gauge 
coupling unification of strong and weak interactions, 
and predicts the correct value for 
$\sin^2\theta_W=3/8$ at the unification scale $M_{\rm GUT}$.

Besides the hypercharge combination, there are two additional 
$U(1)$'s. It is easy to check that one of the two can be identified 
with $B-L$.  For instance, in model A choosing the signs 
$\varepsilon_d=\varepsilon_L=-\varepsilon_\nu=
-\varepsilon_H=\varepsilon_{H'}$, it is given by:
\be
B-L=-{1\over 6}Q_3+{1\over 2}Q_2-{\varepsilon_d\over 2}Q_1\, .
\label{BL}
\ee
Finally, the above spectrum can be easily implemented 
with a Higgs sector, since the Higgs field $H$ has the same 
quantum numbers as the lepton doublet or its complex conjugate:
\ba
&& {\rm Model\ 
A}\qquad\qquad\quad\quad {\rm Model\ B, C}
\nonumber\\
\!\!\!\!\!\!\!\! H\ && ({\bf 1},{\bf 
2};0,-1,\varepsilon_H)_{-1/2}\quad\ ({\bf 1},{\bf 2};0,\varepsilon_H,1)_{-1/2}
\\
\!\!\!\!\!\!\!\! H' && ({\bf 
1},{\bf 2};0,1,\varepsilon_{H'})_{1/2}\qquad\ \, 
({\bf 1},{\bf 
2};0,\varepsilon_{H'},-1)_{1/2}
\nonumber
\label{higgs}
\ea

\section{Mass scales}

\subsection{String scale}

To preserve gauge coupling unification, the compactification scale 
(actually the smallest, if there are several) must be of order of the 
unification scale $M_{\rm GUT}\simeq 10^{16}$ GeV. 
Above this energy, gauge interactions
acquire a higher dimensional behavior.
Moreover, to keep the theory weakly coupled, the string
scale $M_s\equiv{\alpha'}^{-1/2}$ should be close to the 
compactification scale and therefore to $M_{\rm GUT}$.
On the other hand, as we discussed above, 
to ensure that corrections to the unification of gauge couplings
are within 1\%, the magnetic fields should be weak,
$|H_I^a|\alpha'\simlt 0.1$. From the quantization condition
(\ref{Hquant}), it follows that the string scale should be roughly 
a factor of 3 higher than the compactification scale, 
\be
M_s\simeq 3\, M_{\rm GUT}\, .
\label{Ms}
\ee

\subsection{Scalar masses}

The supersymmetry breaking scale $m_0$ is given by the
heaviest charged scalar mass (\ref{scalars}):
$m_0^2\sim\delta H^a\equiv\sum_{I=1}^3 \epsilon_I H_I^a$
on brane stacks, and $m_0^2\sim\delta H^a-\delta H^b$
on brane intersections. Here, $\epsilon_I$ are signs: two
positive and one negative.
Thus, even for strong magnetic fields, of order
of the string scale, $m_0$ can be much smaller and corresponds
to an arbitrary parameter. Although values much lower than 
$M_{\rm GUT}$ require an apparent fine tuning of radii, such a 
tuning is technically natural since the supersymmetric 
point $m_0=0$ is radiatively stable.

All scalar masses are of the order of the supersymmetry breaking scale $m_0$,
which is assumed to be very high in split supersymmetry, except for
those coming from supersymmetric sectors, which are vanishing to 
lowest order, such as the higgses.
The latter are expected to acquire masses from one loop 
corrections, proportional to $m_0$ but suppressed by a loop factor. 
Note that off diagonal elements of the $2\times 2$ Higgs 
mass matrix, usually denoted by $B\mu$, should also 
be generated at the same order as the diagonal elements, 
in the absence of a Peccei-Quinn (PQ) symmetry. For 
high $m_0$, a fine tuning between $B\mu$ and the diagonal 
elements is then required to ensure a light Higgs. 

\subsection{Gaugino masses}

It remains to discuss the corrections to gaugino and 
higgsino masses, $m_{1/2}$ and $\mu$, which are vanishing at the 
tree-level. In the absence of gravity, they are both protected by an 
R-symmetry. Actually, higgsino masses are protected in addition by a 
PQ symmetry which must be broken in order to generate a $B\mu$ 
mixing term in the Higgs mass matrix, as we argued above. 
Then, a $\mu$-term can be generated via $B\mu$, 
or directly using the PQ symmetry breaking, 
if R-symmetry is broken. Indeed,
R-symmetry is in general broken in the gravitational sector 
by the gravitino mass $m_{3/2}$ and thus, in the presence 
of gravity, $m_{1/2}$ and $\mu$ are not anymore protected.
Since supersymmetry breaking in the gravity sector is model
dependent and brings more uncertainties, here we will
assume that gravitational corrections are negligible. 
For instance, if supergravity breaking occurs via a Scherk-Schwarz
compactification on an interval transverse to our 
braneworld~\cite{Antoniadis:1998ki}, using
the usual $\mathbb{Z}_2$ fermion number in the bulk, the gravitino
acquires Dirac mass together with its Kaluza-Klein modes and
R-symmetry remains unbroken~\cite{Antoniadis:2004dt}. 
One can therefore discuss other
sources of R-symmetry breaking within only global supersymmetry.

As discussed previously, supersymmmetry breaking via internal 
magnetic fields is described in the 4d effective field theory by vacuum 
expectation values (VEVs) of D-term auxiliaries for all magnetic $U(1)$'s. 
In the low energy limit, one has:
\be
\langle D\rangle\simeq m_0^2\, ,
\label{Dterm}
\ee
and thus R-symmetry remains unbroken. However, it is broken by
$\alpha'$-string corrections, that modify for instance the gauge
kinetic terms to the DBI form. In particular, gaugino masses can be
induced by a dimension-seven effective operator which is the
chiral F-term~\cite{Antoniadis:2005sd}:
\be
F_{(0,3)}\int d^2\theta{\cal W}^2{\rm Tr}W^2\quad\Rightarrow\quad
m_{1/2}\sim{m_0^4\over M_s^3}\, ,
\label{mgaugino}
\ee
where ${\cal W}$ and $W$ denote the magnetic $U(1)$ and non-abelian
gauge superfield, respectively. The coefficient $F_{(0,3)}$ is a moduli 
dependent function given by the topological partition function on a 
world-sheet with no handles and three boundaries. It is non-vanishing
when the three brane stacks associated to the boundaries do not intersect
at a point in any of the three internal torii. From the effective field theory 
point of view, it corresponds to a two-loop correction involving massive open 
string states. Upon a VEV $\langle{\cal W}\rangle=\theta\langle D\rangle$,
the above F-term generates gaugino masses given in eq.~(\ref{mgaugino}).
They are in the TeV region for scalar masses at intermediate energies,
$m_0\sim{\cal O}(10^{13})$ GeV.

\section{Split extended supersymmetry}

Implementing split supersymmetry in string theory faces a generic
problem: in simple brane constructions the gauge sector comes in
multiplets of extended supersymmetry~\cite{Blumenhagen:2005mu,marc}.
${\cal N}=4$ in the toroidal case, or ${\cal N}=2$ is simple orbifolds. Gauginos can
therefore get Dirac masses without breaking R-symmetry. Indeed, a
Dirac mass~\cite{Fox:2002bu} is induced through the
dimension-five operator
\begin{equation}
\frac{a}{M_s} 
\int d^2\theta {\mathcal W}\, W^a A_a\ \Rightarrow
m_{D}\sim
a \frac{m_0^2}{M_s}\, ,
\label{Dgaugino}
\end{equation}
where $a$ accounts for a possible loop factor.  Actually, this
operator arises quite generally at one-loop level in intersecting
D-brane models with a moduli-dependent coupling, determined only 
from the massless (topological) sector of the theory~\cite{marc}.
Note that this mass $m_D$ is much higher that the Majorana induced
mass of eq.~(\ref{mgaugino}).

It turns out that this scenario is compatible with one-loop gauge coupling
unification~\cite{Antoniadis:2005em}. In the energy regime
between $M_{\rm GUT}$ and the electroweak scale $M_W$, the
renormalization group equations meet three thresholds. From $M_{\rm GUT}$
to the common scalar mass $m_0$ all charged states contribute.  Below
$m_0$ squarks and sleptons (which do not affect unification), adjoint
scalars and $2-n_H$ higgses decouple, while below $m_{D}$ the
${\cal N}=2$ or ${\cal N}=4$ gluinos and winos drop out.  Finally, at TeV energies
higgsinos decouple and we are left with the Standard Model with $n_H$ 
Higgs doublets. Using $M_s\sim M_{\rm GUT}$ and varying $a$
between $a=1$ and $a=1/100$, one finds realistic values for 
$M_{\rm GUT}$ and $m_0$ in both ${\cal N}=4$ and ${\cal N}=2$ cases.  
The results are summarized in Table~\ref{table}.\\
\begin{table}[h]
\begin{tabular}{|c|c|c|c|c|c|}
\hline
 &$n_H$ & $M_{\rm GUT}$  & $m_0$ &$m_{D}$ & $m_{1/2}$ \\
\hline
${\cal N}=2$  & $1$ & $10^{18}$ & $10^{13}$ & $10^6$ & $10^{-5}$ \\
& $2$ & $10^{16}$ & $10^{13}$ & $10^9$ & $10^2$ \\
\hline\hline
 ${\cal N}=4 $& $1$ & $10^{19}$  & $10^{16}$ & $10^{12}$ & $10^6$ \\
 & $2$ & --- & --- & --- & --- \\
\hline
\end{tabular}
\caption{\it Values for the unification scale $M_{\rm GUT}$, scalar masses
$m_0$, Dirac gaugino masses $m_{D}$, and Majorana gaugino masses
$m_{1/2}$ in GeV for ${\cal N}=2,4$
supersymmetric gauge sector and $n_H=1,2$ light higgses.}
\label{table}
\end{table}
In all cases the unification scale is high enough to avoid
problems with proton decay. For the two possible cases with one light
Higgs (${\cal N}=2$ or ${\cal N}=4$), $M_{\rm GUT}$ is very close to the Planck scale so
that there should be no need to explain the usual mismatch between
these two scales.

The low energy sector of these models contains, besides the SM, just
some fermion doublets (higgsinos) and eventually two singlets (the
binos from the discussion below). It therefore illustrates the fact that 
only these states are needed for a minimal extension of the SM consistent 
with unification and Dark Matter (DM) candidates, and not the full fermion 
spectrum of split supersymmetry~\cite{Arkani-Hamed:2005yv}.

Another constraint on the
models is that they must provide a DM candidate.  As usually in
supersymmetric theories this should be the lightest neutralino.  Pure
higgsinos cannot be DM candidates because their mass is of Dirac type.
Since DM direct detection experiments have
ruled out Dirac fermions up to masses of order $50$ TeV, some mixing
coming from the binos is required in order to break the degeneracies
of the two lightest neutralinos; the required mass difference is bigger
than about 150 keV~\cite{Smith:2001hy,Giudice:2004tc,Antoniadis:2005em}. 
This translates into
an upper bound on the Dirac gaugino mass of about $10^5$ GeV, for the
required higgsino mass splitting to be generated through the electroweak
symmetry breaking mixing, which is of order $m_W^2/m_{D}$.
This value compared to the values in Table~\ref{table} leads to
the ${\cal N}=2$, $n_H=1$ case as the only possibility to accommodate
it (the direct Majorana component of bino $m_{1/2}$ is negligible 
in this case).  In fact, one needs an order of magnitude suppression of the 
induced Dirac mass for binos relative to the other gauginos, which is not 
unreasonable to assume in brane constructions.

In the other two models, the required suppression factor is much
higher and the above mechanism would be very unnatural.  However,
since binos play no role for unification as they carry no SM charge,
we could imagine a scenario where $m_{D}$ vanishes identically
for binos, but not for the other gauginos.  For instance consider the
case where Dirac masses from the operator (\ref{Dgaugino}) are
generated by loop diagrams involving ${\cal N}=2$ hypermultiplets with
supersymmetric masses of order $M_{\rm GUT}$ and a supersymmetry 
breaking splitting of order $m_0$.  It is then possible to choose these
massive states such that they carry no hypercharge, in which case
binos can only have Majorana masses displayed in Table~\ref{table}.
Their value in the ${\cal N}=4$ case corresponds to the upper bound for 
DM with Majorana bino mass~\cite{Antoniadis:2005em}.
Thus, the constraint of a viable DM
candidate leaves us with two possibilities: (a) ${\cal N}=2$ with $n_H=1$ and
Dirac masses for all gauginos and (b) ${\cal N}=4$ with $n_H=1$, or ${\cal N}=2$
with $n_H=2$ and Majorana mass binos.

Finally, the higgsinos must acquire a mass of order the electroweak 
scale.  This can be induced by the following dimension-seven operator, 
generated at one loop level~\cite{Antoniadis:2005sd,marc}:
\begin{equation}
\frac{c}{M_s^3}\int d^2\theta \mathcal{W}^2 \overline{D}^2
{\bar H}_1 {\bar H}_2\Rightarrow
\mu \sim c \frac{m_0^4}{M_s^3}\, ,
\label{mu}
\end{equation}
where $c$ is again a loop factor. The resulting numerical value is of
the same order as $m_{1/2}$. Thus, such an
operator can only give a sensible value of $\mu$ for the ${\cal N}=2$ $n_H=2$
model. In the other two cases, ${\cal N}=4$ or ${\cal N}=2$ with $n_H=1$, $\mu$
remains an independent parameter.

To summarize, at low energies we end up with two distinct scenarios
after all massive particles are decoupled: (i) $n_H=1$ with light higgsinos 
(models with ${\cal N}=2$ and ${\cal N}=4$ gauge sector and $n_H=1$), and
(ii) $n_H=2$ with light higgsinos and binos (model with ${\cal N}=2$ gauge
sector and $n_H=2$). In the $n_H=1$ scenario the DM candidate is
mainly higgsino, although the much heavier bino is light enough to
forbid any vector couplings. The relic density reproduces
the actual WMAP results for $\mu\sim 1.1$ TeV.


\begin{theacknowledgments}
This work was supported in part by the European Commission under the
RTN contract MRTN-CT-2004-503369, and in part by the INTAS
contract 03-51-6346.
\end{theacknowledgments}



\bibliographystyle{aipproc}   


\end{document}